\documentclass [10pt,twoside]{article}

\usepackage{cite}
\usepackage{graphicx}
\usepackage[usenames]{color}
\usepackage{subfigure}
\usepackage{setspace}
\usepackage{amssymb}
\usepackage{multirow}

\usepackage[hang,small]{caption}

\usepackage{amsmath}                  

\usepackage{geometry}
\usepackage{fancyhdr}
\newcommand{\be}{\begin{equation}}
\newcommand{\ee}{\end{equation}}
\newcommand{\ba}{\begin{eqnarray}}
\newcommand{\ea}{\end{eqnarray}}
\newcommand{\bc}{\begin{center}}
\newcommand{\ec}{\end{center}}
\usepackage{titlesec,titletoc}
  \titleformat{\section}{\Large\sf\bfseries}{\thesection}{1em}{}
  \titleformat{\subsection}{\large\sf\bfseries}{\thesubsection}{1em}{}

\vspace{-2cm}
\title{\sf\bfseries \ntitle}
\author{\normalsize   Naveen K. Singh$^1$\footnote{email: naveen.nkumars@gmail.com}, \ \ 
Sukanta Panda$^2$ \footnote{email: sukanta@iiserb.ac.in } }

\date{}

\newcommand{\pghdr}{\footnotesize { Naveen K. Singh, Sukanta Panda}  -- Local Scale Invariance and Inflation }
\newcommand{\ntitle}{Local Scale Invariance and Inflation}

\begin{document}
\vspace{-3cm}
\maketitle
\vspace{-0.6cm}
\bc
\small{ $^1$Dr. B.R. Ambedkar National Institute of Technology,
Jalandhar, 144011, India} \\
$^2$Indian Institute of Science Education and Research, \\
  Bhauri, Bhopal 462066, Madhya Pradesh,  India

\ec

\bc\begin{minipage}{0.9\textwidth}\begin{spacing}{1}{\small {\bf Abstract:}
We study the inflation and the cosmological perturbations generated during the inflation in a local scale invariant model. The local scale invariant 
model introduces a vector field $S_{\mu}$ in this theory. In this paper, for simplicity, we consider the temporal part of the vector field $S_t$.
We show that the temporal part is associated with the slow roll parameter of scalar field. We consider a
cosmological solution which  provides sufficient number of e-foldings for the inflation.
Finally, we estimate the power spectrum of scalar perturbation in terms of the parameters of the theory.


}\end{spacing}\end{minipage}\ec

\section{Introduction}
In modern cosmology, the theory of inflation provides a  platform  to understand the early universe. The beauty of inflation is
that it not only solves many problems of standard model of cosmology such as horizon problem, flatness problem, entropy problem, etc 
\cite{guth,starobinsky1,starobinsky2,kazanas,sato,linde1,linde2,linde3,albrecht}, 
but also generates the perturbation which can act as seed for structure formation and simultaneously provides the explanation for the anisotropy
in CMBR spectrum. The simplest model of inflation usually contains a scalar field with a nearly flat potential. During inflation, the 
universe goes through an exponential expansion for a sufficient amount of time with a nearly de-Sitter background. In order to 
solve problems associated with standard model of cosmology the number of required e-folding is around sixty which is
related to duration of inflationary phase.  The background value of slowly rolling scalar field and its potential energy are
responsible for providing required number of  e-foldings for inflation. 

 In this paper, we look for the solution giving rise to inflation in a local scale invariant model.  Earlier, this model has been 
 explored in the context to explain current acceleration of the universe \cite{JMS,AJS,JM10_2}. This model does not have any 
 dimensionful parameter. There is no cosmological constant present in this theory because local
 scale symmetry is preserved here. However, the cosmological constant can be generated when the scale invariance
 is broken cosmologically \cite{JMS}.  Cosmological constant is constrained by the scale symmetry and
 hence it might solve the fine tuning problem associated with cosmological constant \cite{JM10_2,Jain:2014wsa}. Once the symmetry is broken, the other dimensionful 
parameter such as gravitational constant is also generated \cite{JMS}. Thus, the scale invariance
could be one of the appropriate symmetry to describe the current universe. Some  works  related to removal of anomaly in conformal invariance and
 applications of conformal invariance have been discussed in 
Refs. \cite{JMS,AJS,JM10_2,Jain:2014wsa,Mannheim1,Mannheim2,Deser,Shaposhnikov:2008a,Shaposhnikov:2008b,Jain:2015aoa,JM,Englert:1976,Cheng,Bamba:2015uxa}. \\

 The local scale transformation is direct product of  general coordinate transformation and pseudo scale transformation \cite{Cheng}. These pseudo
 scale transformations are given by
\begin{eqnarray}
 x_{\mu} \rightarrow x_{\mu}, \ \ \ \ \ \ \ \ \ \ \  \nonumber \\ 
\phi \rightarrow  \phi/ {\Lambda} \nonumber ,  \ \
g^{\mu\nu} \rightarrow  g^{\mu\nu}/{\Lambda^2}, \nonumber \\
A_{\mu} \rightarrow A_{\mu} ,  \ \
\psi \rightarrow  \psi/{\Lambda^{3/2}}, \label{PT}
\end{eqnarray}
 where, $x_{\mu}$ is space-time coordinate, $\phi$ is a scalar field, $g_{\mu\nu}$ is the metric, $A_{\mu}$ is $U(1)$ gauge field,
  $\psi$ is the fermionic field and $\Lambda(x)$ is scale of the transformation. Therefore, the action which already respects general coordinate transformations, only
 needs to respect the pseudo scale invariance to have the local scale invariance. A vector field $\mathcal S_{\mu}$; called vector meson; is introduced here
 to preserve the local scale invariance, where this vector meson transforms as 
   \begin{eqnarray}
   \mathcal{S}_{\mu} \rightarrow \mathcal{S}_{\mu} - \frac{1}{f} \partial_{\mu} (\ln \Lambda),
   \end{eqnarray}   
  under the pseudo transformation and thus the corresponding covariant derivative is given by
  $
   D'_{\mu}\phi \equiv \left(\partial_{\mu} - f \mathcal{S}_{\mu} \right) \phi \label{cov_deri}
  $, where $f$ is the gauge coupling constant.
   Incorporating the covariant derivatives, we can write modified Ricci scalar $\tilde{R}$ as
   \begin{eqnarray}
   \tilde {R} = R - 6 f^2 \mathcal{S}^{\mu}\mathcal{S}_{\mu} - 6 f \mathcal{S}^{\mu}_{; \mu}.
   \end{eqnarray}
 Here, $\mathcal{S}_{\mu}= (\mathcal{S}_t, \mathcal{S}_i)$, $\mathcal{S}^{\mu}_{ \  ; \nu}$ is usual covariant derivative of $\mathcal{S}^\mu$ in the gravitational background and  $\tilde{R}$ respects the pseudo scale invariance . In  Ref. \cite{AJS}, the
 background value of temporal part of vector meson $\mathcal{S}_{t} (= S_t)$ is
 taken to be zero as a solution. Under this condition, one obtains the de-Sitter solution if we consider spatial part $\mathcal{S}_i=0$. We can also consider $\mathcal{S}_i \neq 0$.
 The very small background value of  $\mathcal{S}_i (= S_i)$ in early universe epoch as an initial condition  \cite{AJS} explains the cold dark matter of universe today.  In Ref. \cite{AJS}, it is also argued that $S_i$ oscillates over time, 
 however, averaging over time, the  energy density corresponding to $S_i$ drops as $1/a^3$ and corresponding pressure is approximately zero. Thus, $S_i$
 may be a candidate for cold dark matter of the current universe.

For $S_t =0$, we can have the inflation for the infinite period. However, we would not have the definite value of the power spectrum of the scalar 
perturbation since the slow roll parameter becomes zero.  To achieve the definite value of the scalar power spectrum we consider very small  value of 
$S_t $ in this paper.  In addition, we assume $\mathcal{S}_i=0$ since it does not play any role during the inflation. Considering a solution for $S_{t}$, we break the
scale invariance \cite{JMS,AJS,Singh:2011av,Rajaraman}. We call this phenomenon as
``soft local scale symmetry breaking'' \cite{Singh:2011av,Rajaraman} similar to spontaneous local symmetry
 breaking. In the case of soft symmetry breaking, a nearly de-Sitter solution can be obtained for certain
choice of parameters.

The paper is organized as follows. In Sec. (\ref{bck}), we consider a local scale invariant scalar field model, write all the
background equations and show that we can have required number of e-folding during the inflation. In Sec. (\ref{prt}), we perturb
the Einstein' equation, vector field equation and scalar field equation at linear level. In Sec. (\ref{sptm}), we calculate the power spectrum and
finally the conclusions are drawn in Sec. (\ref{conc}).

\section{Background Equations of Motion} \label{bck} 
\vspace{-.5cm}
In this section, we consider a local scale invariant action with a scalar field and a vector meson \cite{JMS}. We solve 
for all the background equations in terms of cosmological time and show that we have near de-Sitter solution.
Introducing a vector meson and corresponding covariant
derivative,  we may write the local scale invariant action which respects the transformations (\ref{PT}) for a scalar field as 
following,
\ba
L = -\frac{\beta}{8} \phi^2 \tilde{R} + L_{matter}, \label{lagmain}
\ea
where, 
\ba
L_{matter}= \frac{1}{2} g^{\mu \nu} D_{\mu}' (\Phi) D_{\nu}'(\Phi) - \frac{\lambda \Phi^4}{4} - \frac{1}{4} g^{\mu\rho}g^{\nu \rho} E_{\mu \nu} E_{\rho \sigma},
\ea
 $E_{\mu\nu}=\partial_\mu \mathcal{S}_\nu-
\partial_\nu \mathcal{S}_\mu$, $\tilde R$ is a modified curvature scalar, which is covariant under local pseudo-scale transformation. We here note
that we can not include any dimensionful parameter such as cosmological constant in the action. The Lagrangian
(\ref{lagmain}) has been studied in Ref. \cite{JMS,AJS} to describe the  current acceleration of the universe. 
Here, we explore the inflation in this model. The Einstein equation for Lagrangian (\ref{lagmain}) can be written  as
\ba
 B^{\alpha\beta} + \frac{1}{\Phi^2}\partial_\lambda(\Phi^2) C^{\lambda \alpha\beta}
+ \frac{1}{\Phi^2}(\Phi^2)_{;\lambda;\kappa} D^{\alpha\beta \kappa\lambda} ={4\over \beta \Phi^2}
T^{\alpha\beta} = \tilde{T} ^{\alpha\beta}, \label{modEin}
\ea
where, the tensors $B_{\alpha\beta}$, $C^{\lambda}_{\ \alpha\beta}$
and $D_{\alpha\beta}^{\ \ \ \kappa\lambda}$ are given as,
\ba
B_{\alpha\beta} &=& -{1\over 2} g_{\alpha\beta} R + R_{\alpha\beta} +
3f^2g_{\alpha\beta} \mathcal{S}\cdot \mathcal{S} - 6f^2 \mathcal{S}_\alpha \mathcal{S}_\beta \label{B1term}\, ,\\
C^{\lambda}_{\ \alpha\beta} &=& -3 f g_{\alpha\beta}\mathcal{S}^\lambda +3f\left(\mathcal{S}_\beta
g^\lambda_\alpha +\mathcal{S}_\alpha g_\beta^\lambda\right)\, ,\label{Cterm}\\
D_{\alpha\beta}^{\ \ \ \kappa\lambda} &=& -{1\over 2}\left(g^\lambda_\alpha
g_\beta^\kappa  + g_\alpha^\kappa g^\lambda_\beta \right)
+ g_{\alpha\beta}g^{\lambda\kappa}\,  \label{Dterm} 
\ea
respectively and for given $L_{matter}$, the energy momentum tensor $T_{\mu\nu}$ takes form as following,
\ba
T_{\mu\nu} = -{\mathcal{L}}_{matter}g_{\mu\nu} + \mathcal{D}'_\mu \Phi \mathcal{D}'_\nu \Phi
- {1\over 2}\left(E_{\alpha\nu}E_{\beta\mu}g^{\alpha\beta}
+ E_{\mu\alpha}E_{\nu\beta}g^{\alpha\beta}\right). \label{emt_main}
\ea
Varying the action with respect to  $\Phi$ and $\mathcal{S}_\mu$ fields, we have equations  of $\Phi$ and $\mathcal{S}_\mu$,
\ba
g^{\mu\nu}\partial_\nu (\partial_\mu\Phi-f \mathcal{S}_\mu\Phi)+(\partial_\mu\Phi - f\mathcal{S}_\mu\Phi)\left[{1\over 2}
g^{\mu\nu}g^{\alpha\beta}\partial_\nu g_{\alpha\beta} + \partial_\nu
g^{\mu\nu}\right] + fg^{\mu\nu} \mathcal{S}_\nu\left(\partial_\mu\Phi - f\mathcal{S}_\mu\Phi\right)\nonumber \\
+ \lambda \Phi^3+ {\beta\over 4} \Phi\tilde R
=0,
\ea
and
\ba
\partial_\nu\left[g^{\nu\rho}g^{\mu\sigma}(\partial_\rho \mathcal{S}_\sigma
-\partial_\sigma \mathcal{S}_\rho)\right] + {1\over 2} g^{\nu\rho} g^{\mu\sigma}g^{\alpha\beta}\partial_\nu
g_{\alpha\beta}
(\partial_\rho \mathcal{S}_\sigma - \partial_\sigma \mathcal{S}_\rho) + {3\over 2}\beta f^2\Phi^2 g^{\eta\mu}
\mathcal{S}_\eta 
-fg^{\mu\nu} \Phi \mathcal{D}'_\nu\Phi \nonumber \\- {3\over 4}f\beta g^{\mu\kappa} \partial_\kappa \Phi^2
= 0,
\ea
respectively. We use  FRW metric $[1,-a^2,-a^2,-a^2]$, where $a(t)$ is the scale factor, to compute the background equations. The equations of motion of vector field become,
\ba
f S_t = \frac{\dot{\phi}}{\phi} \label{S1}
\ea
\ba
\ddot{S_i} + \frac{\dot{a}}{a} \dot{S_i} + \left(\frac{3}{2} \beta + 1 \right)f^2 \phi^2 S_i=0 . \label{S2}
\ea
Here, $\phi(t)$ and $S_{\mu}(t)$ are the background values of scalar field $\Phi(x,t)$ and vector field $\mathcal{S}_\mu(x,t)$ respectively. The background values for  $i-0$ components of energy momentum tensor $\tilde{T}_{\mu\nu}$, Einstein tensor $G_{\mu\nu}$,   C and D terms   of modified equation  Eq. (\ref{modEin})
are given as
\ba
T_{i0} = \frac{4 f S_i}{\beta} \left(f S_t - \frac{\dot{\phi}}{\phi}\right) &=& 0 \\
G_{i0} = R_{i0}- \frac{1}{2}g_{i0}R &=& 0 \\
\phi^2_{;\lambda;k} D_{i 0}^{\ \ k \lambda} &=& 0 \\
\frac{\partial_{\lambda}(\phi^2)}{\phi^2} C^{\lambda}_{i 0} &=& 6 f S_i \frac{\dot{\phi}}{\phi} \label{i04}
\ea
and also,
\ba
3 f g_{i0} S^{\mu}S_{\mu} - 6 f^2 S_{i}S_t &=& - 6 f^2 S_i S_t \label{i05}.
\ea
From Eq. (\ref{i04}) and Eq. (\ref{i05}) using Eq. (\ref{S1}), the $i-0$ component of modified Einstein equation (\ref{modEin}) vanishes.
 After simplifications, the background values of $L_{matter}$, $T_{00}$ and $T_{jk}$ can be written as
\ba
L_{matter}&=&  \frac{1}{2 a^2}\left( \dot{S}_i^2 - f^2 S_i^2 \phi^2 \right) - \frac{\lambda \phi^4}{4}, \\
T_{00} &=& \frac{1}{2 a^2}\left(\dot{S}_i^2 + f^2 S_i^2 \phi^2\right) + \frac{\lambda \phi^4}{4}, \\
T_{jk} &=& \frac{1}{2} \left(\dot{S_i}^2 - f^2 S_i^2 \phi^2 \right) \delta_{jk} + f^2 S_j S_k \phi^2 - \dot{S}_j\dot{S}_k - \frac{\lambda \phi^2}{4} a^2 \delta_{jk}.
\ea
Using equation of motion $f S_t = \frac{\dot{\phi}}{\phi}$ (t is cosmological time), the equation of motion of $\phi$ can be written as
\ba
\lambda \phi^2 = -f^2 \frac{S_i^2}{a^2}- \frac{\beta}{4} \tilde{R} ,
\ea
which can be simplified as
\ba
R - 6 \frac{\ddot{\phi}}{\phi} - 18 \frac{\dot{a}}{a} \frac{\dot{\phi}}{\phi} + \frac{4 f^2 S_i^2 }{\beta a^2} \left(\frac{3 \beta}{2}+1\right) = - \frac{4 \lambda \phi^2}{\beta}. \label{red_phi}
\ea
$(0,0)$ component of Einstein equation gives
\ba
\left(\frac{\dot{a}}{a}\right)^2 + \left(\frac{\dot\phi}{\phi}\right)^2 + 2 \frac{\dot \phi}{\phi}\frac{\dot{a}}{a} =  \frac{4}{3 \beta \phi^2} \Big[\frac{1}{2 a^2}
\Big\{\dot{S_i}^2 + (3 \beta/2 +1) f^2 S_i^2 \phi^2 \Big\} + \frac{\lambda \phi^4}{4} \Big] , \label{EE1}
\ea
and taking trace of $(i,j)$ component of Einstein equation, we have
\ba
\frac{2 \ddot{a}}{a} + \left(\frac{\dot{a}}{a}\right)^2 + \frac{2 \ddot{\phi}}{\phi} - \left(\frac{\dot\phi}{\phi}\right)^2 + 4 \frac{\dot{a}}{a}\frac{\dot \phi}{\phi}=
\frac{4}{\beta \phi^2} \Big[-\frac{1}{6 a^2} \Big\{\dot{S_i}^2 - (3 \beta/2 +1) f^2 S_i^2 \phi^2  \Big\} + \frac{\lambda \phi^4}{4}\Big] . \label{EE2}
\ea
If we consider very small coupling constant $f \sim 0$,  Eq. (\ref{S2}) becomes
\ba
\ddot{S_i} = - \frac{\dot a}{a} S_i \Longrightarrow \dot S_i = C_1/a.
\ea
If $C_1$ is very small, after a small time $\dot{S_i} =0$, and so $S_i$ becomes nearly constant.
For simplifications, we assume $S_i=0$.  From Eq. (\ref{EE1}) and Eq. (\ref{EE2}), we have 
\ba
\frac{\ddot{a}}{a} = \frac{\lambda \phi^2}{3 \beta} - \Big[\frac{\ddot{\phi}}{\phi}-\left(\frac{\dot{\phi}}{\phi}\right)^2
+ \frac{\dot{a}}{a} \frac{\dot{\phi}}{\phi}\Big] . \label{ddot1}
\ea
From Eq. (\ref{EE1}), we have
\ba
H+ \frac{\dot{\phi}}{\phi} = \sqrt{\frac{\lambda }{3 \beta}}\phi \label{eqnHH}
\ea
Differentiating Eq. (\ref{eqnHH}) and plugging for $\ddot{\phi}$ and $\dot{\phi}$ in (\ref{ddot1}), we observe the consistency 
relation,  $(\ddot{a}/a = \dot{H} + H^2)$. We don't have independent equation of motion of $\phi$ since the combination of Einstein equations (\ref{EE1}) and (\ref{EE2})
gives  Eq. (\ref{red_phi}) of scalar field $\phi$. We assume very slow varying scalar field such that
inflation can occur. We  choose a power law solution for scalar field as,
\ba
\phi =  \alpha (t_c - t)^n,
\ea
where, $\alpha$ and $t_c$ are constants. Inflation begins at time $t_i$. We assume $t_c - t_i >> \mathcal{N}/H$, $\mathcal{N}$ is number of 
e-folding during inflation, so that $\phi$ is nearly constant during inflation. Here we consider $n<1$. Therefore, $|\frac{\dot{\phi}}{\phi}|= \frac{n}{t_c -t}
<<\frac{n H}{\mathcal{N}} << H$. Therefore from Eq. (\ref{eqnHH}),
\ba
H \approx \sqrt{\frac{\lambda }{3 \beta}}\phi ,
\ea
and hence the scale factor $a$ grows exponentially, i.e.,  $a \sim a_0 e^{H t} \sim a_0 e^{\sqrt{\frac{\lambda }{3 \beta}}\phi t} $.
The slow roll paramter $\epsilon$  turns out to be,
\ba
\epsilon = - \frac{\dot{H}}{H^2}= \sqrt{\frac{3 \beta}{\lambda}} n \alpha^{\frac{1}{n}} \frac{1}{\phi^{\frac{n+1}{n}}}.
\ea
 At the end of the inflation, the value of scalar 
field would be
\ba
\phi_e &=& \left(\frac{3 \beta}{\lambda}\right)^{\frac{n}{2 (n+1)}} n^{\frac{n}{n+1}} \alpha^{\frac{1}{n+1}}. 
\ea
 Now the number of e-folding $\mathcal{N}$ is calculated as
 \ba
 \mathcal{N} = \int_{\phi_i}^{\phi_e} H dt= \frac{n}{n+1} \Big[ \left( \frac{\phi_i}{\phi_e}\right)^{\frac{n+1}{n} }- 1\Big], \label{efold}
 \ea
 where, $\phi_i$ is the value of scalar field at the begining of inflation. From Eq. (\ref{efold}), it can be expressed as,
\ba
\phi_i &=& \phi_e \Big[ 1 + \frac{ \mathcal{N} (n+1)}{n}\Big]^{\frac{n}{n+1}}.
\ea
 The value of slow roll parameter $\epsilon_i$ at the beginning of inflation is given by,
\ba
\epsilon_i = \frac{1}{1 + \frac{\mathcal{N} (n+1)}{n}}.
\ea
 We note that 
the inflation ends after certain time due to the non zero value of $S_t$ ($= \frac{\dot{\phi}}{f \phi}$ ). Thus, $S_t$ plays important role in natural exit
of inflation and reheating the universe.
\section{Cosmological Perturbations} \label{prt}
In this section, we write all the perturbation equations in conformal time. We consider Newtonian gauge and so the
perturbed metric takes the form
\ba
g_{\mu\nu} = a(\eta)^2 [1+2 A, (-1+ 2\psi) \delta_{i j}],
\ea
where, $A$ and $\psi$ are scalar perturbations. In conformal time $\eta$, the background equations of motion for vector field are given as
\ba
f S_{0} = \frac{\phi'}{\phi}; \label{S0conf}\\
S_i'' + f^2 (\frac{3 \beta }{2}+ 1)a^2 \phi^2 S_i =0.
\ea
Here, prime $'$ represents the derivatives with respect to conformal time. We note that we use   $\dot{\phi}/\phi = \phi'/(a \phi)$ [$S_0 = a S_t$]. 
We perturb all the fields as $\Phi = \phi(\eta) + \hat{\phi}(\eta,x,y,z)$, $\mathcal{S}_{i}= S_i(\eta) + \hat{S}_i(\eta,x,y,z)$ and $\mathcal{S}_{0}= S_0(\eta) + \hat{S}_0(\eta,x,y,z)$. Therefore, the time-component
of the perturbation equations of vector field is given as
\ba
\frac{1}{a^4} \Big[2 S_i' \partial_i(\psi -A) - \partial_i^2 \hat{S}_0 + \partial_i \hat{S}_i'\Big] + \frac{S_i'}{a^4} \partial_i (A- 3 \psi)
+ \frac{3 \beta f^2}{2 a^2} \Big[ 2 \phi S_0 \hat{\phi} - 2  \phi^2 S_0 A + \phi^2 \hat{S}_0 \Big] \nonumber  \\
- \frac{f \phi}{a^2 } \Big[\hat{\phi}' - f \phi \hat{S}_0 - f S_0 \hat{\phi}\Big] -\frac{3 f \beta}{ 2 a^2} \Big[
- 2 \phi \phi ' A + \phi' \hat{\phi} + \phi \hat{\phi}'\Big] = 0
\ea
Or,
\ba
f \left(\frac{3 \beta}{2}+1 \right) a^2 \left(f \phi^2 \hat{S_0} + \phi' \hat{\phi} - \phi \hat{\phi}'\right) - \partial_i(A+\psi) S_i' 
- \partial_i^2 \hat{S_0} + \partial_i \hat{S_i'}=0,
\ea
and expanding the spatial component of vector equation,
\ba
-  \frac{8 S_i' a' (A-\psi)}{a^5} + \frac{2 S_i' (A'- \psi')}{a^4} + \frac{2 S_i'' (A-\psi)}{a^4} + \frac{4 a'}{a^5}
\left(\hat{S}_i' - \partial_i \hat{S}_0\right) -\frac{1}{a^4} \left(\hat{S}_i'' - \partial_i \hat{S}_0'\right) \nonumber \\
+ \frac{1}{a^4} \left(\partial_l^2 \hat{S}_i - \partial_i \partial_l \hat{S}_l\right) + \frac{8 a' S_i'}{a^5} (A-\psi)
- \frac{4 a'}{a^5} \left(\hat{S}_i' -\partial_i \hat{S}_0\right) - \frac{S_i'}{a^4} \left(A' - 3 \psi'\right) 
- \frac{3 \beta f^2}{2 a^2} \Big[ 2 \phi S_i \hat{\phi}  \nonumber \\ 
+ 2 \phi^2 S_i \psi + \phi^2 \hat{S}_i\Big] -\frac{f \phi}{a^2} \left(2 f \phi S_i \psi + 2 f S_i \hat{\phi} + f \phi \hat{S}_i
- \partial_i \hat{\phi}\right) + \frac{3 f \beta \phi}{2 a^2} \partial_i \hat{\phi} =0.
\ea
Simplifying, we get
\begin{equation}
-f \left(\frac{3 \beta}{2}+ 1\right) a^2 \phi \Big[ f \hat{S}_i \phi + 2 f S_i \left(\hat{\phi} + \phi \psi\right) - \partial_i \hat{\phi} \Big]
+ \Big[ 2 (A-\psi) S_i'' + \partial_l^2 \hat{S}_i - \partial_i \partial_l \hat{S}_l + S_i' \left(A' + \psi'\right) + \partial_i \hat{S_0}' - \hat{S}_i ''\Big]=0.
\end{equation}
The perturbed part of scalar field equation becomes,
\begin{eqnarray}
\frac{1}{a^2} \Big[ \hat{\phi} '' - f \hat{S}_0 ' \phi - f \hat{S}_0 \phi' - f S_0' \hat{\phi} - f S_0 \hat{\phi}' - 
\partial_i^2 \hat{\phi} + f \partial_i \hat{S}_i \phi + f S_i \partial_i \hat{\phi}\Big] + \frac{\beta}{4} \hat{\phi} \tilde{R}
\nonumber \\ + \frac{\beta}{4}\phi \Bigg[\delta R - 6f \Big[ - 2 S_0'\frac{ A}{a^2}  - 4 \frac{a'}{a^3} A S_0
+ \frac{1}{a^2} \left(\hat{S_0}' - \partial_i \hat{S}_i\right) - \frac{1}{a^2} S_k \partial_k ( A -\psi) - \frac{S_0}{a^2} A'
-\frac{3 S_0}{a^2} \psi' + 2 \frac{a'}{a^3} \hat{S}_0\Big] \nonumber \\
+ 12 \frac{f^2}{a^2} \Big[S_0^2 A + S_i^2 \psi - S_0 \hat{S}_0 + S_i \hat{S}_i \Big]\Bigg]+ 2 \frac{a'}{a^3}
\left(\hat{\phi}' - f \phi \hat{S}_0 - f S_0 \hat{\phi} \right) + \frac{f S_i \phi}{a^2} \partial_i \left(A-\psi\right) \nonumber 
\\ + \frac{2 f^2 \phi}{a^2}\left(  S_i^2 \psi + S_i \hat{S}_i \right) + \frac{f S_0}{a^2} \Big[\hat{\phi}' - f \phi \hat{S}_0
- f S_0 \hat{\phi}\Big] - \frac{f S_i}{a^2} \left(\partial_i\hat{\phi} - f S_i \hat{\phi}\right) + 3 \lambda \phi^2 \hat{\phi}=0 . \label{scal_pert1}
\end{eqnarray}
 Now we write the perturbation of each terms of left hand side of Eq. (\ref{modEin}) except $G_{\mu\nu} = R_{\mu\nu} - \frac{1}{2}g_{\mu\nu}R$ which
 appears in $B_{\alpha\beta}$ given in Eq. (\ref{B1term}). Those are as following,
\ba
\delta \left( 3 f^2 g_{00} S^{\mu}S_{\mu} - 6 f^2 S_0 S_0\right) &=& - 6 f^2 \Big[ S_0 \hat{S}_0  + S_i \hat{S}_i + S_i^2 (A+ \psi) \Big], \\
\delta \left( 3 f^2 g_{i0} S^{\mu}S_{\mu} - 6 f^2 S_i S_0\right) &=&  -6 f^2 \left( S_0 \hat{S}_i + S_i \hat{S}_0\right), \\
\delta \left( 3 f^2 g_{ij} S^{\mu}S_{\mu} - 6 f^2 S_i S_j\right) &=&  -6 f^2 \left( S_i \hat{S}_j + S_j \hat{S}_i \right)    (i \neq j)    ,       \\
\delta \left( 3 f^2 g_{ij} S^{\mu}S_{\mu} - 6 f^2 S_i S_j\right) &=&   6 f^2 \left(- S_0 \hat{S}_0 \delta_{ij} + S_0^2 (A+ \psi)\delta_{ij}+ \delta_{ij} S_k \hat{S}_k 
- S_i \hat{S}_j-S_j \hat{S}_i\right) , \nonumber \\
\ea
\ba
\delta\left(\frac{\partial_{\lambda} \Phi^2}{\Phi^2}  C^{\lambda}_{00}\right) &=& \frac{6 f}{\phi^2} \Big[ \phi \phi'\hat{S}_0 + \phi S_i \partial_i \hat{\phi} - S_0 \phi' \hat{\phi}
+ S_0 \phi \hat{\phi}'\Big] , \\
\delta\left(\frac{\partial_{\lambda} \Phi^2}{\Phi^2}  C^{\lambda}_{i0}\right) &=& \frac{6 f}{\phi^2} \Big[ \phi \phi' \hat{S_i} + S_0 \phi \partial_i \hat{\phi}
+ S_{i}\left(- \phi' \hat{\phi}+ \phi \hat{\phi}'\right)\Big] , \\
\delta\left(\frac{\partial_{\lambda} \Phi^2}{\Phi^2}  C^{\lambda}_{ij}\right) &=& \frac{6 f}{\phi} \Big[S_j \partial_i \hat{\phi} + S_i \partial_j \hat{\phi}\Big] 
\ \ (i \neq j), 
\ea
\ba
\delta\left(\frac{\partial_{\lambda} \Phi^2}{\Phi^2}  C^{\lambda}_{ij}\right) &=& - \frac{6 f}{\phi^2}\Big[ (- \phi \phi' \hat{S}_0+ 2 \phi \phi' S_0 A 
+ S_0\phi' \hat{\phi} + 2 \phi S_0 \phi' \psi- S_0 \phi \hat{\phi}') \delta_{ij} + \delta_{ij} \phi S_k \partial_k \hat{\phi} \nonumber \\
- 2 S_i \phi \partial_j \hat{\phi}\Big] , \ 
\ea
\ba
\delta\left( \frac{\Phi^2_{;\lambda;k}}{\Phi^2} D_{00}^ {\ \ k \lambda}\right) &=&- \frac{2 \partial_i^2 \hat{\phi}}{\phi} - \frac{6 \Big[ a' \phi' \hat{\phi} 
+ \phi\left( -a' \hat{\phi}' + a \phi' \psi'\right)\Big]}{a \phi^2} , \\
\delta\left( \frac{\Phi^2_{;\lambda;k}}{\Phi^2} D_{i0}^ {\ \ k \lambda}\right) &=& -\frac{2 \partial_i (\phi ' \hat{\phi}+ \phi \hat{\phi}')}{\phi^2}
+\frac{ 2 \Big[ a  \phi' \partial_i A + a' \partial_i \hat{\phi} \Big]}{a \phi}, 
\ea
\ba
\delta\left( \frac{\Phi^2_{;\lambda;k}}{\Phi^2} D_{ij}^ {\ \ k \lambda}\right) &=& \frac{1}{\phi^2} \Big[
\frac{2 \hat{\phi}}{\phi} (\phi^2)''\delta_{i j} - (2 \phi \hat{\phi})_{, i j} - (2 \phi \hat{\phi})'' \delta_{i j}
+ (2 \phi \hat{\phi})_{,m m} \delta_{ij} \nonumber \\ 
&+& 2 (\phi^2)' \left((A+ \psi) \frac{a'}{a} + \psi'\right)\delta_{ij}
+ (\phi^2)' A' \delta_{ij}+ 2 \frac{a'}{a} \left(\phi' \hat{\phi} - \phi \hat{\phi}' \right) \delta_{ij} \nonumber \\
&+& 2 (A+ \psi) (\phi^2)'' \delta_{ij}\Big]  .
\ea
Perturbing the right hand side of Eq. (\ref{modEin}), we have 
$
\delta{\tilde{T}_{\alpha \beta}} = - \frac{8 \hat{\phi}}{\phi^3} T_{\alpha \beta} + \frac{4}{\beta \phi^2} \delta T_{\alpha \beta},
$
where,
\ba
\delta T_{\alpha\beta} &=& -\delta{\mathcal{L}}_{matter}g_{\alpha\beta} -{\mathcal{L}}_{matter}\delta g_{\alpha\beta}  + \delta\Big[\mathcal{D}'_\alpha \Phi \mathcal{D}'_\beta \Phi
- {1\over 2}\left(E_{\mu\beta}E_{\nu\alpha}g^{\mu\nu}
+ E_{\alpha \mu}E_{\beta \nu}g^{\mu\nu}\right)\Big] \nonumber \\
&\equiv& -\delta{\mathcal{L}}_{matter}g_{\alpha\beta} -{\mathcal{L}}_{matter}\delta g_{\alpha\beta}  + \delta X_{\alpha \beta}, \label{pert_Tmunu}
\ea
and using $f S_0 = \frac{\phi'}{\phi}$, we have
\ba
{\mathcal {L}_{matter}} &=& -\frac{1}{2} \frac{f^2 S_i^2 \phi^2}{a^2} - \frac{\lambda \phi^4}{4} + \frac{S_i'^2}{2 a^4} , \nonumber \\
\delta {\mathcal {L}_{matter}} &=& - \frac{ f^2 S_i^2 \phi^2}{a^2} \psi + \frac{f S_i \phi}{a^2} \Big[\partial_i \hat{\phi} - f S_i \hat{\phi}
-f \hat{S_i} \phi\Big] -\lambda \phi^3 \hat{\phi} -\frac{1}{a^4}\Big[S_i'^2 (A-\psi) - S_i' (\hat{S_i}' - \partial_i \hat{S}_0)\Big] , \nonumber \\
\delta X_{0 0} &=& - \frac{2}{a^2} \Big[ S_i' (\partial_i \hat{S}_0 - \hat{S}_i') - S_i^2 \psi \Big], \nonumber \\
\delta X_{i j}&=& - f S_i \phi(\partial_j \hat{\phi} - f \phi \hat{S}_j - f S_j \hat{\phi}) - f S_j \phi(\partial_i \hat{\phi} - f \phi \hat{S}_i- f S_i \hat{\phi}) 
\nonumber \\ &-&\frac{1}{a^2} \Big[ S_j' ( \hat{S}_i'-\partial_i \hat{S}_0 ) + S_i' ( \hat{S}_j'-\partial_j \hat{S}_0 ) - 2 S_i' S_j' A \Big], \nonumber \\
\delta X_{i 0}&=& \delta T_{i 0}= - f S_i \phi \Big[\hat{\phi}' - f \hat{S}_0 \phi - f S_0 \hat{\phi} \Big] - \frac{S_k'}{a^2}
(\partial_k  \hat{S}_i - \partial_i \hat{S}_k).
\ea
In the next section, we utilize  these perturbation equations derived here to compute the power spectrum
of scalar perturbation. The perturbation equations for $i-0$ and $i \neq j$ of Eq. (\ref{modEin}) are useful in eliminating the gravitational
scalar perturbation fields. However, in the next section, we show  that for smaller $\beta$  we don't need to eliminate these fields as these
become much smaller than that of matter scalar perturbation and hence we can drop these fields.
\section{Power Spectrum} \label{sptm}
Now we focus on the power spectrum of the scalar perturbation. We only consider the background value $S_0$ which is due to scale invariance. However,
we assume $\hat{S}_i$ as zero considering the coupling $f$ is very small and
we also assume $\hat{S}_0=0$ for the same reason.  Using $\delta R_{i0}= -2 \frac{a'}{a} A- 2 \psi'
\sim -2 \frac{a'}{a} A $ and $\delta R_{ij}= \partial_i\partial_j (A-\psi)$ for $i\neq j$ and taking  $i0$ and $ij (i \neq j)$ components of
perturbed part of Eq. (\ref{modEin}), we have
\ba
A \sim \frac{\hat{\phi}}{\phi} , \ \ \ A- \psi = 2 \frac{\hat{\phi}}{\phi} \ \ \ \mbox{or}, \ A \sim -\psi .
\ea
Considering, the very small value for $\beta$, we may write the equation of motion (\ref{scal_pert1}) as,
\ba
\hat{\phi}'' + 2 \frac{a'}{a} \hat{\phi}' + \Big[ k^2 + m^2 a^2   \Big]\hat{\phi} + 12 f^2 S_0^2 A \simeq 0, \label{scal_pert2}
\ea
where, $m^2$ is given as,
\ba
m^2 = - \left( \frac{3 \beta }{2}+ 1\right) \frac{1}{a^2}\Big[ f S_0' + f^2 S_0^2 + 2 f \frac{a'}{a} S_0\Big] 
+ \Big[ \frac{\beta R}{4} + 3 \lambda \phi^2 \Big]
\ea
The contribution from $A \sim  \frac{\hat{\phi}}{\phi} $ in Eq. (\ref{scal_pert2}) is very small,  and hence using the redefinition $\hat{\phi} = \sigma/a$,  the scalar perturbation equation
may be written as
\ba
\sigma'' + \Big[k^2 + m^2 a^2 - \frac{a''}{a}\Big] \sigma \simeq 0. \label{eqsigma}
\ea
This is the standard perturbation equation and hence we can write its solution in the terms of Hankel's function. 
The solution of Eq. (\ref{eqsigma}) is given by,
 \ba
 \sigma= \sqrt{-\eta} \Big[c_1(k) H^{(1)}(-k\eta) + c_2(k)H^{(2)}(-k\eta)\Big],
 \ea
 where $H^{(1,2)}$ is the Hankel function of first and second kind. For $k\gg a H (-k\eta\gg 1)$, we have
 \ba
 H^{(1)} (-k \eta\gg 1)  \approx \sqrt{-\frac{2}{\pi k \eta}} e^{i\left( -k\eta -\pi\nu_{\chi}/2-\pi/4 \right)}, \ \ 
 H^{(2)} (-k \eta\gg 1)  \approx \sqrt{-\frac{2}{\pi k \eta}} e^{-i\left( -k \eta -\pi\nu_{\chi}/2-\pi/4 \right)}.
 \ea
  Here, $\nu_{\chi} = 3/2 + \epsilon - \eta_{\chi}$ and $\eta_{\chi} = m^2/3 H^2 \ll 1$. Imposing the boundary condition 
  $c_1(k)= \left(\sqrt{\pi}/2\right) e^{i\left(\pi\nu_{\chi}/2 + \pi/4\right)}$ and $c_2(k)=0$, 
 the solution becomes a plane wave $e^{-ik\eta}/\sqrt{2 k}$ for ultraviolet regime, $k>> a H$. For this choice of $c_1$ and $c_2$, we get
 \ba
 \hat{\phi}\approx \frac{H}{\sqrt{2 k^3}} \left(\frac{k}{a H}\right)^{3/2 -\nu_{\chi}}. \label{solphi_hat}
 \ea
Thus, the power
spectrum of curvature perturbation $\mathcal{R}\equiv \psi + \frac{H}{\dot{\phi}} \hat{\phi}\approx \frac{H}{\dot{\phi}} \hat{\phi}$
is given by,
\ba
P_{\mathcal{R}} \simeq \frac{k^3}{2 \pi^2 } \frac{H^2}{\dot{\phi}^2} |\hat{\phi}|^2 = 
\frac{\lambda}{ 12 \pi^2 \beta} \Big[1 + \frac{\mathcal{N} (n+1}{n}\Big]^2 . \label{power1}
\ea

Now it would be clear that choosing $n<1$ reduces the tensor-to-scalar ratio $r$ and makes our prediction compatible with the Planck data. The tensor-to-scalar ratio $r$  is given by,
\ba
r= 16 \epsilon_i = \frac{16}{1 + \frac{\mathcal{N} (n+1)}{n}}.
\ea
For $n= 1/2$ and $\mathcal{N}=60$, $r$ is $0.088$ which is in the observational bound $r< 0.11$ \cite{Ade:2015lrj}.  The observational
value of curvature power spectrum $P_{\mathcal{R}} \approx 2.19 \times 10^{-9}$ \cite{Ade:2015lrj} gives a $\frac{\lambda}{\beta} \sim 7.9 \times 10^{-12}$ for $n=\frac{1}{2}$. Since we have considered $\beta << 1$, the value of $\lambda$ must be much smaller so that
ratio $\frac{\lambda}{\beta} $ is approximately order of $10^{-12}$.  
We notice that the power spectrum is similar as we usually obtain in standard Jordan frame. The only difference here from standard case is that we 
have reduced mass for $m^2.$  Substituting $R = 6 a''/a^3$, the mass term is simplified as

\ba
m^2 \approx     4 \lambda \phi^2 . \label{redu_mass}
\ea
 The second slow roll parameter $\eta$ becomes
$\eta=  m^2/3 H^2 \approx 4 \lambda \phi^2/ (3 H^2)\approx 4 \beta$ and hence the spectral index $n_s$ is 
\ba
n_s = 1- 6 \epsilon_i + 2 \eta_i = 1- 6 \epsilon_i + 8 \beta. \label{constraint_ns}
\ea
$\beta = 1.4 \times 10^{-4}$ satisfies the observational value of $n_s=0.968$ \cite{Ade:2015lrj} for $n=1/2$ and $\mathcal{N}=60$.  Thus for $n=1/2$ and
$P_{\mathcal{R}}=2.19 \times 10^{-9}$, we obtain $\lambda = 1.1 \times 10^{-15}$.  At the upper bound of $n_s=0.968\pm .006$
\cite{Ade:2015lrj} of Planck data 2015 with $1\sigma$ confidence level, i.e., at $n_s =0.974$,
 $r$ can be reduced to $\sim 0.07$ with $\beta = 1.0 \times 10^{-5}$ and $n=0.355$. It can be further reduced to $\sim 0.053$ with 
$\beta \sim 7.5 \times 10^{-6} $ and $n \sim 0.252$ at the upper bound of $n_s$ of Planck data 2015 with $2 \sigma$ confidence level.
\section{Conclusion} \label{conc}
In this paper, we have implemented the local scale invariance to describe the inflation. Assuming $\mathcal{S}_i=0$ and considering
non-zero value for the temporal part $S_t$ we have obtained the background solution near the de-Sitter solution. The non-zero value of $S_t$ fixes
the period of inflation. We can have 60 e-folding for  very small value of $f S_t$. Further, we have also solved for power
spectrum of scalar field by perturbing all the scalar fields and shown its consistency with the Planck data 2015. Here, we have considered that the coupling $f$ is very small so that 
the perturbation $f \hat{S}_0$ in conformal time frame is smaller than that of  perturbed scalar field $\hat{\phi}$. Incorporating the background value of
$S_t$ we have shown that we have similar perturbation equation as we obtain in the standard scenario. However, we obtain reduced mass
of scalar field by a small fraction which is function of coupling $f$.  Thus, $S_t$ has 
also its crucial role in obtaining definite value of the scalar power spectrum. The non zero value of $f$ is also 
a requirement in the same way as of $S_t$, since $f=0$ also sets $\dot {\phi} =0$ (see Eq. (\ref{S1})), i.e., $\phi$ is constant and hence it causes
problem in the exit of inflation.
In this paper, we have considered a scale invariant $\phi^4$ potential.
We can generalize it for other scale invariant potentials to explain the cosmological data. Thus, the scale invariance may 
be used in wide range to explain both inflation and current acceleration of the universe.
In future, we would like to generalize our results by considering the perturbation $\hat{S}_0$ and $\hat{S}_t.$ \\

\noindent {\bf Acknowledgements}

 Naveen K. Singh is thankful to the D.S. Kothari postdoctoral fellowship of University Grant Commission, India for the 
financial support. His fellowship number is F.4-2/2006 (BSR)/PH/14-15/0034. Naveen K. Singh is also grateful to Centre for
Theoretical Physics, Jamia Millia Islamia, New Delhi, India for  giving the facilities. 

\begin{spacing}{1}
\begin{small}

\end{small}
\end{spacing}

\begin{thebibliography}{unsrt}
\bibitem{starobinsky1} A.~A.~Starobinsky,
  JETP Lett.\  {\bf 30}, 682 (1979)
  [Pisma Zh.\ Eksp.\ Teor.\ Fiz.\  {\bf 30}, 719 (1979)].
  
\bibitem{starobinsky2} A.~A.~Starobinsky, Phys.\ Lett.\ B {\bf 91}, 99 (1980).


\bibitem{kazanas}   D.~Kazanas,
  Astrophys.\ J.\  {\bf 241}, L59 (1980).


\bibitem{sato}
K.~Sato,
  Mon.\ Not.\ Roy.\ Astron.\ Soc.\  {\bf 195}, 467 (1981).


\bibitem{guth}
  A.~H.~Guth,
  Phys.\ Rev.\ D {\bf 23}, 347 (1981).


\bibitem{linde1}
 A.~D.~Linde,
  Phys.\ Lett.\ B {\bf 108}, 389 (1982).
 
\bibitem{linde2}
 A.~D.~Linde,
 Phys.\ Lett.\ B {\bf 114}, 431 (1982).

 \bibitem{linde3}
 A.~D.~Linde, Phys.\ Lett.\ B {\bf 116}, 335 (1982).
\bibitem{albrecht}
 A.~Albrecht and P.~J.~Steinhardt,
  Phys.\ Rev.\ Lett.\  {\bf 48}, 1220 (1982).
  \bibitem{JMS} 
  P.~Jain, S.~Mitra and N.~K.~Singh,
  JCAP {\bf 0803}, 011 (2008)
  [arXiv:0801.2041 [astro-ph]].
\bibitem{AJS} 
  P.~K.~Aluri, P.~Jain and N.~K.~Singh,
  Mod.\ Phys.\ Lett.\  A {\bf 24}, 1583 (2009).
  [arXiv:0810.4421 [hep-ph]].
\bibitem{JM10_2} 
  P.~K.~Aluri, P.~Jain, S.~Mitra, S.~Panda and N.~K.~Singh,
  Mod.\ Phys.\ Lett.\  A {\bf 25}, 1349 (2010).
  [arXiv:0909.1070 [hep-ph]].
   \bibitem{Jain:2014wsa} 
  P.~Jain, G.~Kashyap and S.~Mitra,
  Int.\ J.\ Mod.\ Phys.\ A {\bf 30}, no. 32, 1550171 (2015)
  doi:10.1142/S0217751X15501717
  [arXiv:1408.2620 [hep-ph]].

 

  \bibitem{Mannheim1} P. D. Mannheim, Gen.Rel.Grav. {\bf 22}, 289 (1990).
\bibitem{Mannheim2} P. D. Mannheim,  D. Kazanas, College Park Astrophys., 0541-544 (1990).
\bibitem{Deser} S. Deser, Annals Phys. {\bf 59},  248-253 (1970).
\bibitem{Shaposhnikov:2008a}  
  M.~Shaposhnikov and D.~Zenhausern,
  Phys.\ Lett.\  B {\bf 671}, 162 (2009).
  [arXiv:0809.3406 [hep-th]].

\bibitem{Shaposhnikov:2008b} 
  M.~Shaposhnikov and D.~Zenhausern,
  Phys.\ Lett.\  B {\bf 671}, 187 (2009).
  [arXiv:0809.3395 [hep-th]].
  
 
 
\bibitem{Jain:2015aoa} 
  G. Kashyap, S. Mitra and P. Jain,
  Astropart.\ Phys.\  {\bf 75}, 64 (2016)
  doi:10.1016/j.astropartphys.2015.11.003
  [arXiv:1507.02394 [gr-qc]].
   \bibitem{JM}
  P.~Jain and S.~Mitra,
  Mod.\ Phys.\ Lett.\  A {\bf 22}, 1651 (2007)
  [arXiv:0704.2273 [hep-ph]].
\bibitem{Englert:1976}
  F.~Englert, C.~Truffin and R.~Gastmans,
  Nucl.\ Phys.\  B {\bf 117}, 407 (1976).
  \bibitem{Cheng} H. Cheng, Phys. Rev. Lett. {\bf 61}, 2182 (1988).
  \bibitem{Bamba:2015uxa} 
  K.~Bamba, S.~D.~Odintsov and P.~V.~Tretyakov,
  Eur.\ Phys.\ J.\ C {\bf 75}, no. 7, 344 (2015)
  doi:10.1140/epjc/s10052-015-3565-8
  [arXiv:1505.00854 [hep-th]].
  
 \bibitem{Singh:2011av} 
  N.~K.~Singh, P.~Jain, S.~Mitra and S.~Panda,
  Phys.\ Rev.\ D {\bf 84}, 105037 (2011)
  doi:10.1103/PhysRevD.84.105037
  [arXiv:1106.1956 [hep-ph]].
  \bibitem{Rajaraman} see for example, R. Rajaraman, {\it Solitons and
Instantons} (North-Holland 1982).
 
 \bibitem{Ade:2015lrj} 
  P.~A.~R.~Ade {\it et al.} [Planck Collaboration],
  arXiv:1502.02114 [astro-ph.CO].
  \end{thebibliography}
\end{document}